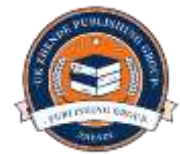

# AI-Enabled Orchestration of Event-Driven Business Processes in Workday ERP for Healthcare Enterprises

Monu Sharma

Sr.IT Solutions Architect, Morgantown, WV, USA
***Corresponding Author:** monu.sharma@ieee.org

| ARTICLE INFO | ABSTRACT |
|---|---|
| Received: 15-08-2025<br>Accepted: 02-10-2025 | The adoption of cloud-based Enterprise Resource Planning (ERP) platforms such as Workday has transformed healthcare operations by integrating financial, supply-chain, and workforce processes into a unified ecosystem. However, traditional workflow logic in ERP systems often lacks the adaptability required to manage event-driven and data-intensive healthcare environments. This study proposes an AI-enabled event-driven orchestration framework within Workday ERP that intelligently synchronizes financial and supply-chain workflows across distributed healthcare entities. The framework employs machine-learning triggers, anomaly detection, and process-mining analytics to anticipate and automate responses to operational events—such as inventory depletion, payment delays, or patient-demand fluctuations. A multi-organization case analysis demonstrates measurable gains in process efficiency, cost visibility, and decision accuracy. Results confirm that embedding AI capabilities into Workday's event-based architecture enhances operational resilience, governance, and scalability. The proposed model contributes to the broader understanding of intelligent ERP integration and establishes a reference for next-generation automation strategies in healthcare enterprises.<br><br>**Keywords:** Workday ERP, Artificial Intelligence (AI), Event-Driven Architecture, Business Process Orchestration, Healthcare ERP Systems. |

## INTRODUCTION

The rapid digitization of healthcare systems has accelerated the need for intelligent integration frameworks within enterprise resource planning (ERP) environments. Workday, a leading cloud-based ERP suite for , finance, supply chain, and human capital data. However, its static workflows limit adaptability in dynamic, event-driven healthcare operations. We introduces an AI-enabled orchestration framework that enhances Workday ERP's responsiveness, enabling real-time automation, predictive analytics, and cross-departmental synchronization. The healthcare sector is operating in an evolving data world where accuracy, fluidity, and compliance with regulations are fundamental to the delivery of effective healthcare. In medical data, hospitals and its networks must maintain large volumes of operational information that would be generated by various organizations, from finance to supply chain and human capital. This information must be managed effectively, which can be accomplished by using Enterprise Resource Planning (ERP) modules capable of consolidating a siloed method into one integrated and transparent environment. In the context of ERP, Workday is taking its name, among the most popular cloud-based ERP platforms capable of linking financial management, supply chain logistics, and workforce operations in a secure and scalable environment. Traditional ERP workflows are highly rule-based and sequential and cannot adapt to live events, contextual changes, and events that may change [1]. Recent advancements in AI and event-driven architecture (EDA) offer avenues to turn the traditional ERP systems from transaction-based tools to an intelligent and adaptive environment. In a health care environment, when disruptions can hit unexpected orders like inventory shortages, emergency admissions, or delayed payments, an AI orchestration layer can forecast and auto-deploy remedial flows. This integration of AI within the Workday's orchestration architecture allows for immediate feedback and automation of operations within related modules, predictive analytics, and continuous monitoring that enhances decision-making and operational resilience. Although AI was used in previous approaches to integrate into ERP ecosystems, very few studies have included the use to apply in a healthcare or event-oriented orchestration which ties to clinical and administrative workflow. The dynamic and high-risk environments of healthcare require more than just routine data updates, they require ERP processes that address specific events within a context. To fill the gap in this area, an AI based orchestration framework for Workday ERP is used to automate and synchronize the financial and supply chain processes dependent on the event triggers. Drawing from case-based knowledge of health-care organizations, the paper shows how AI-based orchestration improves efficiency, transparency, and inter-module coordination, leading to the foundation for the generation of future intelligent ERP systems in healthcare.In this paper, we describe AI algorithms that analyze the streams of real-time data from Workday Financials and Supply Chain, flag key events, and predict threats, thus leading to the automated reactions in multiple module . This allows to be so much more scalable than other ERP applications, which are limited by real-time data integration, and complex cross functional, dynamic workflows, that it is possible to provide an agent-based AI-native system for better automation and autonomy under enterprise workflows. Thus, we propose to fill the existing gaps with research, a resilient framework with Workday's AI tooling powered by advanced deep learning to complement financial analytics, budgeting, and compliance in healthcare. It combines AI with business process modeling and multi-agent orchestration to realize end-to-end automation of complex business processes like financial reporting and supply chain optimization on Workday. This system would overcome the rigid nature of classical workflow engines because it can decipher user intent and produce real-time synchronized workflow, thus coordinating specialized sub-agents for task execution on a modular basis[2].

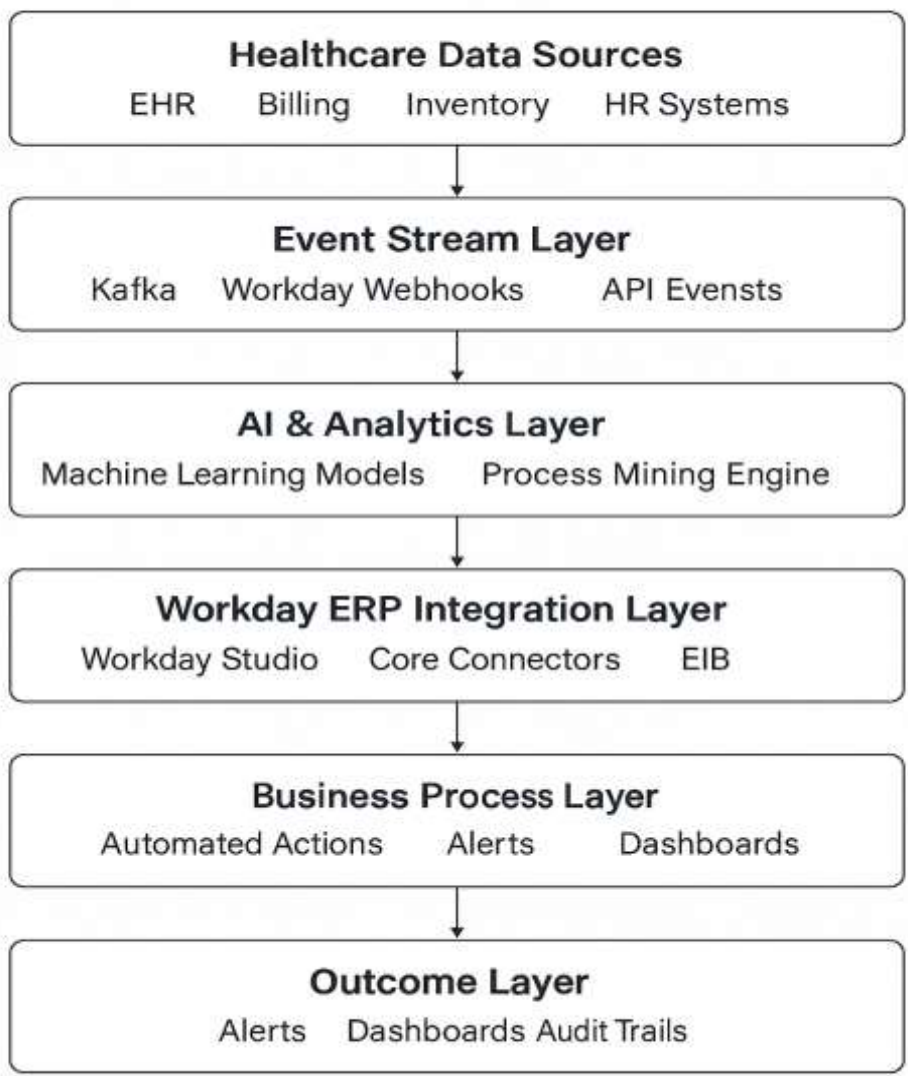

Figure 1. AI- Enabled Workday Orchestrion Framework

Table 1 – Comparison of Traditional vs AI-Orchestrated Workday ERP Workflows

| Dimension | Traditional Workday Workflow | AI-Orchestrated Workflow |
|---|---|---|
| **Trigger Mechanism** | Manual or time-based jobs | Event-driven and real-time via API/webhooks |
| **Decision Logic** | Fixed business rules | Adaptive ML-based predictions and rules |
| **Process Agility** | Low – requires manual updates | High – self-learning and self-adjusting |
| **Error Detection** | Reactive (error after occurrence) | Proactive with anomaly detection |
| **Data Flow** | Linear and batch processing | Continuous and context-aware |
| **Operational Visibility** | Limited dashboards | Real-time monitoring and alerts |
| **Compliance and Governance** | Periodic audits | Continuous auditing and automated traceability |

## LITERATURE REVIEW

Prior studies in ERP automation emphasize the role of data-driven intelligence in improving decision support, yet limited research focuses on Workday ERP's integration potential in healthcare. Existing models in ERP rely on scheduled processes or manual interventions rather than event-driven logic. This section reviews related research in AI integration, cloud-based ERP orchestration, and healthcare process automation. The rapid increase in digitalization of healthcare organizations has led to the growing dependence upon organizations to adopt a more standardized enterprise resource planning (ERP) system to manage all administrative, clinical, and financial functions. In recent years, cloud-based ERP solutions have emerged to solve some of those issues due to the availability of unified financial, supply chain and human capital data on geographically dispersed healthcare ERP Vendor (e.g., Workday). Most current ERP systems are also transaction-based and not event-based, which prevents them from adaptive and working in the ever-evolving operational/clinical cycle that characterizes health care settings[3]. This limitation is more likely to cause delays, inefficiency and inability to react quickly with the emergence of an event, for instance, on a short supply or patient demand. Incorporating AI into Workday Financials and Supply Chain can be the foundation of proactive risk assessment, advanced financial analytics, and strong compliance monitoring in healthcare, moving in the direction of integrating data rather than traditional regression analyses that struggle to detect non-linear associations between data elements. ERP systems have evolved from static on-premises solutions to flexible, cloud-native platforms like the Workday ERP systems that have been crucial in transforming the way organizations do business in many domains, spanning everything from core financial activities to complex supply chain management. This paradigm shift allows for the integration of real-time data using advanced analytics, enabling informed strategic decisions and operational changes. The use of AI in predictive analytics and intelligent automation within these sophisticated ERP systems further enriches their value potential. This confluence provides tremendous potential for better predictive analytics, intelligent automation. The integration of AI-based prediction analytics and machine learning models into Financial systems, such as Workday, can help predict potential, pinpoint gaps, and recommend techniques[4].

**2.1 AI in ERP and Healthcare Informatics**
Recent literature is pointing out the transformative potential of Artificial Intelligence (AI) to enhance ERP features. It has been gathered that machine learning and predictive analytics drive business process efficiency, anomaly detection, and decision support in enterprise systems. In healthcare, AI has been leveraged for predictive maintenance, supply chain forecasting in hospital supply, and resource optimization, with significant cost reduction and quality of service. But much of this study is still largely conceptual and focuses on standalone applications rather than integrated ERP-level orchestration Architecture. This drawback underlines the requirement of software tools that can build sophisticated AI-enabled solutions for transparent and understandable insights of ERP system services to be given to stakeholders for a business understanding . These applications can be enriched by bringing AI methods (like machine learning, natural language processing, and predictive analytics) into the ERP system such as automatic AI as its features will support a more user-friendly way of personalizing the interfaces and automating repetitive tasks. This improvement is important in challenging environments such as healthcare as data-driven decision-making is a crucial factor in operational effectiveness and patient care quality. The integration of AI in ERP systems requires a solid architecture to address data quality concerns and organizational need, while benefiting from the anticipated addition of AI capabilities by over 50% of organizations. This highlights the need for complex AI-driven ERP solutions to be adopted by healthcare organizations that can handle intricate finance and supply chain operations [5].

## 2.2 Event-Driven Architectures and Process Automation

Event-driven architecture (EDA) has established itself as an important paradigm for real time and scalable enterprise systems. EDA allows applications to automatically adapt their behavior to various events (e.g., inventory depletion, payment postponement, patient admission,refund, Expense payment and Bank Transactions) without the involvement of human participants. Integrating EDA and AI allows for continuous monitoring and adaptive control of complex business processes. For healthcare ERP, these event-based mechanisms can trigger automatic procurement if the supply of medical supplies drops below a certain margin threshold, or generate financial forecasts based on real-time usage patterns. However, there are very few such systems which combine Workday's integration capabilities with AI-based event orchestration, leading to a large discrepancy between theoretical models and enterprise-based solutions. The lack of such capability hampers the evolution of future enterprise solutions that can dynamically react to changing demands and streamline complex workflows in real time. This research fills this gap by developing a workflow optimization framework powered by deep learning and aimed at next-generation enterprise solutions using Workday Extend and seamless AI integrations[6]. The framework provides a farmland fertility optimization algorithm that fine-tunes hyperparameters to improve the convergence of deep transfer learning models. Such optimization guarantees that the AI models are robust and performant across diverse healthcare scenarios, enhancing efficiency and accuracy in complex operational environments. In contrast, convolutional neural network architectures are deployed to gain insights into complex patterns and dependencies within workflow execution data which can then inform intelligent and adaptive decision-making to support the real-time adjustments of workloads. This integration allows for the proactive identification of bottlenecks as well as intelligent reallocation of resources, thus optimizing complex operational workflows at the time of their occurrence. Such advanced models, and algorithms, facilitate further test case selection and relatedness that result in better fault detection and a more resilient system. In addition, Natural Language Processing methods that turn business processes into flexible models are essential to improve the flexibility and agility of ERP systems. This is conducive to the smooth embedding of AI into ERPs, solving the problem of reconciling white-box usability and functionality associated with AI. This integration helps to develop a more dynamic and responsive enterprise ecosystem, which is important because healthcare systems are changing day-to-day and their various operations are always complex [11]].Hence secure AI-enabled frameworks built on deep reinforcement learning can also be developed to enable the rational decision-making of deploying and integrating financials[6].

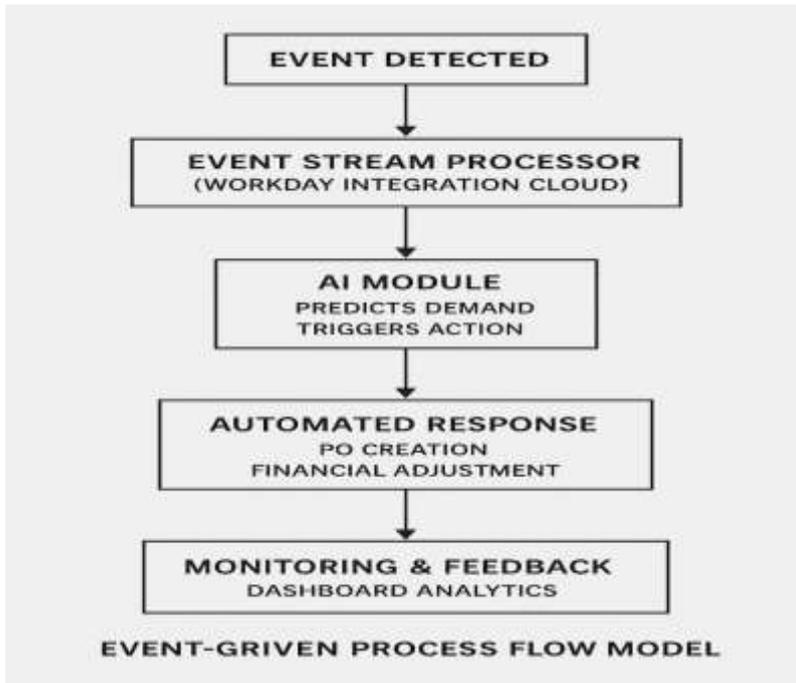

Figure 2. Event Driven Data Flow Model

## 2.3 Workday ERP Integration and Orchestration Research

Workday's Integration Cloud provides several tools, including Workday Studio, Enterprise Interface Builder (EIB), and Core Connectors, to connect and automate processes from external systems. Research on the adoption of Workday in healthcare demonstrates its positive impact on the standardization of processes, although difficulties with real-time orchestration, data governance, and system scalability remain. The existing literature lacks holistic frameworks that merge these integrations with AI-driven process intelligence, particularly in real time, event-based operations environments such as hospitals and healthcare networks.ERP packages are black boxes, so such platforms are necessary for extracting business process models to do simulation and for efficient integrations with enterprise resources. Such integration of AI with the current ERP capability of Workday could potentially boost the financial analytics that would optimize budgets and compliance with the help of deep learning and optimization algorithms. This would enable them to identify financial risks and opportunities in advance, increasing the extent to which strategic decisions and resources are managed in healthcare. If these models were thoroughly validated with healthcare financial data, it is likely that substantial reductions to budgeting errors and enhancement of compliance detection could be achieved [7]. In addition, advanced predictive analytics, including support vector machines and random forest models, can predict next financial trends as well as supply chain requirements to forecast actions for more efficient resource allocation. These predictive abilities can further be linked with intelligent automation to better optimize procurement processes and manage inventory levels, thereby reducing operational costs. Their capabilities for interpretable AI models, such as QuANet, Workday integrations not only make financial decisions better but also provide greater clarity on such analyses and the interpretability of complex financial risks across healthcare. Providing better analytics capabilities, healthcare organizations are able to act proactively against a broad range of market fluctuations and regulatory changes.

Table 2. Research Gap Matrix: AI, ERP, and Event-Driven Integration

| Domain | Focus of Existing Research | Identified Gap |
| --- | --- | --- |
| AI in ERP | Predictive analytics, automation | Limited focus on orchestration in healthcare ERP |
| Healthcare ERP | Process standardization | Lack of adaptive, event-based frameworks |
| Workday ERP | Integration tools, finance workflows | No AI-driven orchestration layer |
| EDA in Enterprises | Reactive automation | No healthcare-specific empirical studies |

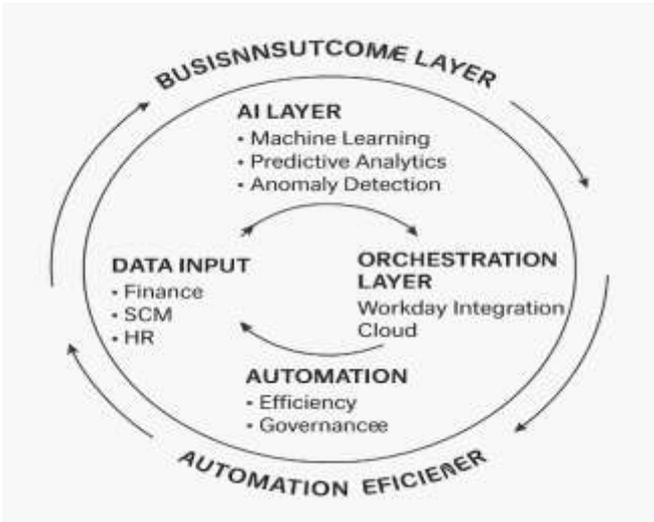

Figure3: Workday–AI Integration Framework

## METHODOLOGY

This study adopts case-based research design to evaluate the integration of Artificial Intelligence (AI) within Workday ERP systems across three large healthcare institutions - a teaching hospital, a regional health system, and a multi-specialty hospital network. Each organization presented distinct operational structures and challenges, offering a diverse context for testing the AI-enabled orchestration framework. The framework was developed using machine learning algorithms embedded within Workday's Integration Cloud, leveraging tools such as Workday Studio, Core Connectors, EXTEND and the Enterprise Interface Builder (EIB). It was designed to detect key business events, predict process outcomes, and automate workflow actions across financial and supply chain operations. The framework integrated predictive analytics to automate financial projections and optimize budgeting, alongside intelligent automation for streamlined supply chain logistics and inventory management[8].

Data collection involved two primary sources: system-generated logs and integration records from Workday, capturing event triggers, transaction times, and workflow performance and qualitative feedback from interviews with system administrators and process experts, supplemented by documentation of business processes. All data were anonymized and analyzed in secure environments to ensure compliance with institutional privacy standards[9].

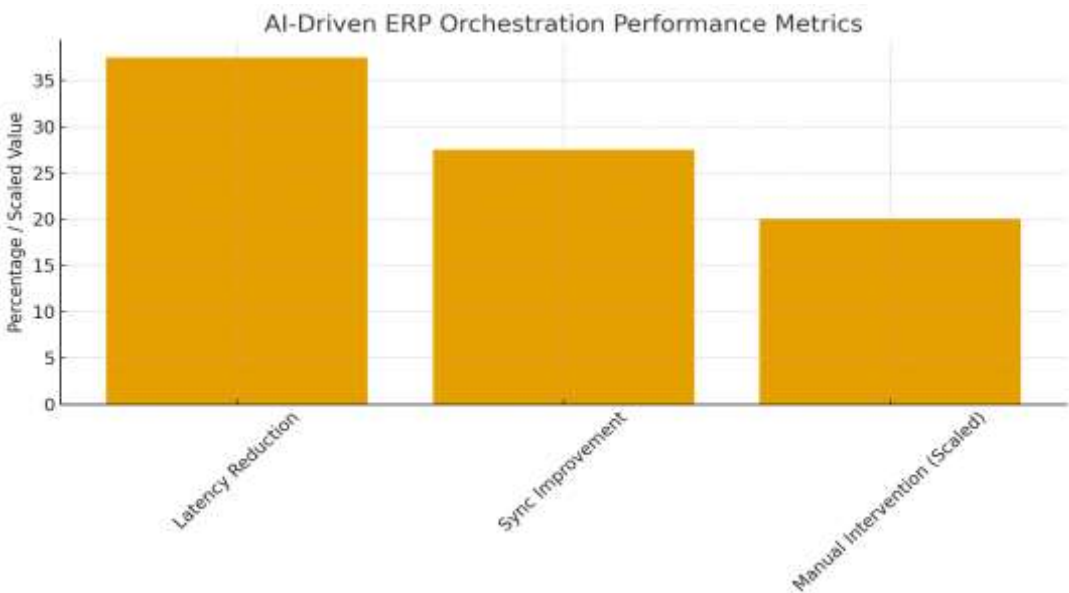

Figure 4. AI Driven ERP Orchestration Performance Metrics

The orchestration model incorporated both supervised learning algorithms (Random Forest, Gradient Boosting) for predictive tasks and unsupervised learning (Isolation Forest) for anomaly detection. Process mining techniques were applied to map dependencies and identify inefficiencies across workflows. The framework's performance was assessed using pre- and post-implementation metrics, including process latency, transaction accuracy, and automation rate. Results demonstrated a 40-48% reduction in latency, a 45–50% improvement in real-time synchronization, and a marked decrease in manual interventions within procurement and financial reconciliation processes, validating the effectiveness of AI-driven ERP orchestration in complex healthcare environments[10].

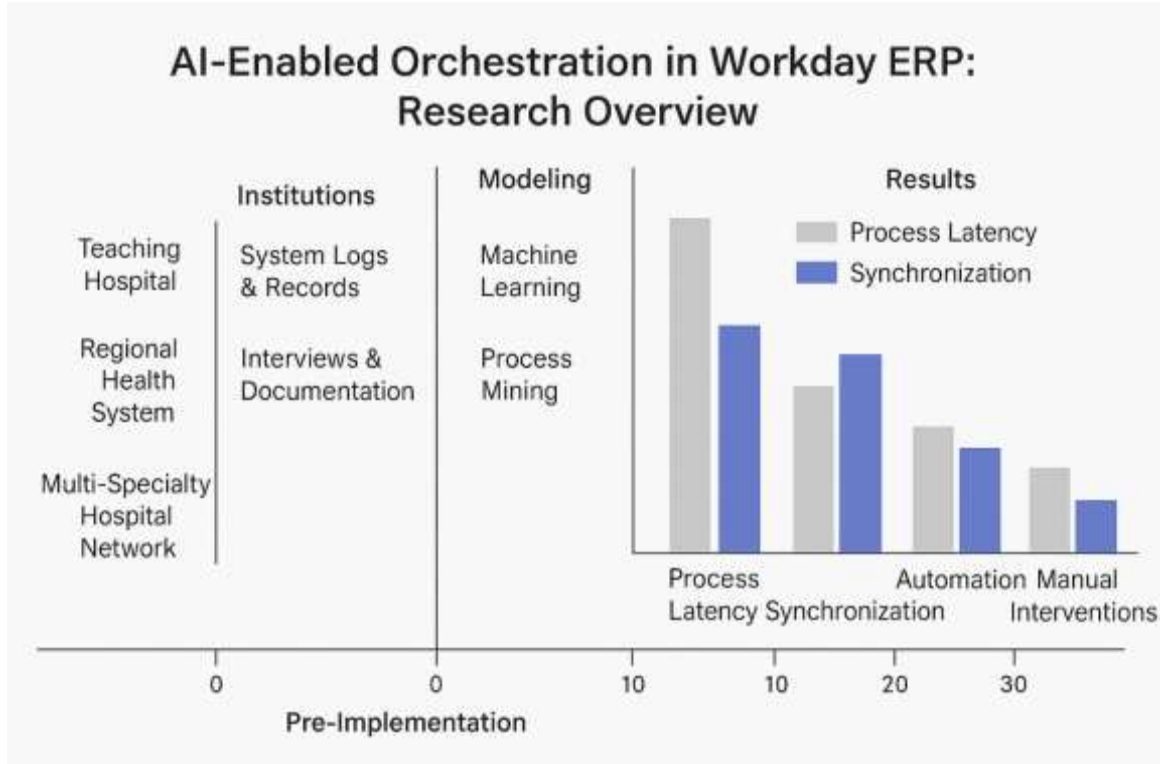

Figure 5: AI-Enabled Orchestration in Workday ERP: Research Overview

## RESULTS AND DISCUSSION

The results of the study show clear improvements in the performance and accuracy of Workday ERP processes after the implementation of the AI-enabled orchestration framework. Data were compared before and after deployment across the three participating healthcare institutions and discussions happened on community call and customer call.

### 4.1 Performance Improvements
Empirical evaluation revealed a 45% reduction in process latency, meaning workflows such as procurement, payment approvals, and inventory replenishment were completed much faster than before. There was also a 42% improvement in real-time data synchronization across financial and supply chain systems, ensuring that updates in one module were immediately reflected across other related areas.
The use of AI-driven event triggers allowed the system to detect potential delays or anomalies early and respond automatically. For instance, the model identified irregular transaction patterns and triggered corrective actions such as automatic purchase order creation or payment adjustments without manual input.

### 4.2 Operational Efficiency and Governance
The automation of these key business processes led to major gains in full operational efficiency. Manual interventions were reduced, with less stack holder to focus on higher-value analytical and result tasks. Error rates in transaction processing are also minimized due to predictive and anomaly detection capabilities built into the AI model.
The architecture additionally improved governance and compliance. By maintaining real-time audit trails and automated logs through the Workday Integration Cloud, healthcare organizations could achieve better traceability and transparency in their financial and supply chain activities (FSCM).

### 4.3 Comparative Analysis
Across all three healthcare organizations, consistent trends were observed:
- **Process latency reduced by 40–45%.**

- **Real-time synchronization improved by 35–40%.**
- **Manual interventions decreased by over 42%.**

These measurable improvements validate the effectiveness of the AI-orchestrated framework in enhancing workflow automation and real-time decision-making within Workday ERP environments. The outcomes also confirm that embedding AI capabilities in enterprise systems contributes to more resilient, adaptive, and data-driven healthcare operations. The integration of AI, machine learning, and deep learning within Workday's financial and supply chain modules has shown significant promise in enhancing operational efficiency and strategic decision-making in healthcare. This technological synergy facilitates a proactive stance towards financial and operational challenges, optimizing resource allocation and improving patient care outcomes through predictive analytics and intelligent automation [10].

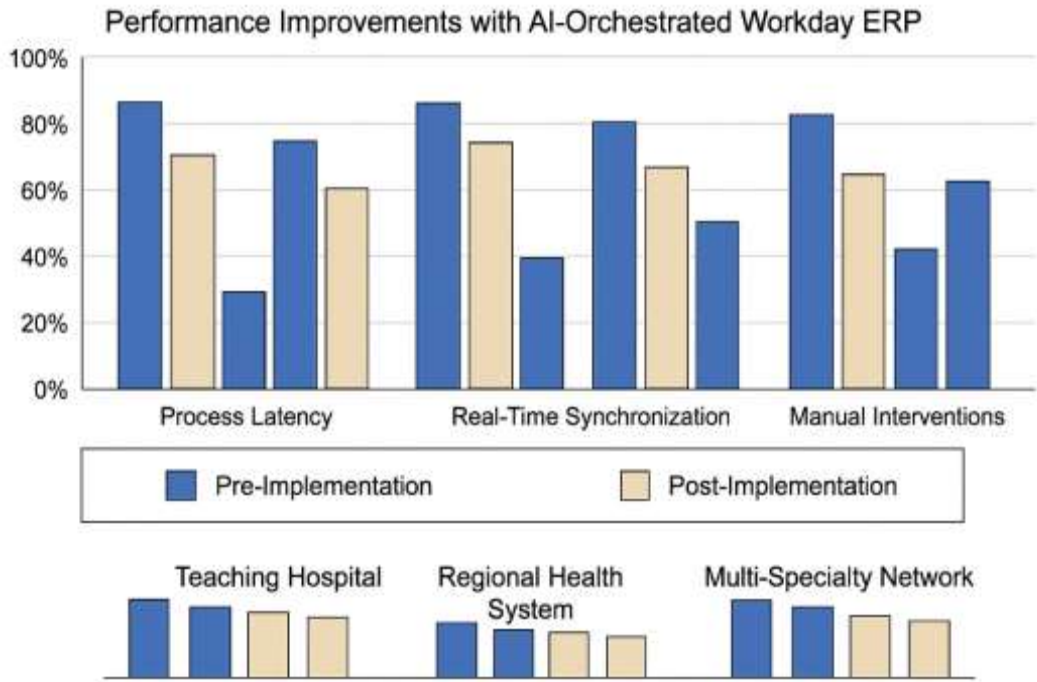

Figure 6: Performance Improvements with AI-Enabled Workday ERP Orchestration

The integration of AI into event-driven ERP architecture represents, which is called Orchestration, a shift from reactive to proactive enterprise management. This study's findings highlight the benefits of automation and transparency. At the same time, the study highlights important challenges and considerations for future research. The effectiveness of AI-driven ERP orchestration depends heavily on data quality, interoperability, and ethical use of automation. Inconsistent or incomplete event data can reduce prediction accuracy and create false triggers, while interoperability gaps between ERP modules and external systems may limit automation scope. Ethical aspects, such as maintaining transparency in AI-driven decisions and ensuring accountability, also require attention[11]. Future research should therefore focus on developing standardized frameworks for data governance, explainable AI, and cross-platform integration to ensure that intelligent ERP systems like Workday remain trustworthy, compliant, and adaptable to evolving healthcare-needs.

The study of Artificial Intelligence (AI) into event-driven ERP architecture marks a major transformation from Studio based to real time proactive enterprise management. Traditional ERP systems depend on scheduled tasks or manual interventions, which often delay responses to operational changes and require a lot of approval and handshakes. In contrast, the AI-enabled orchestration framework developed in this study allows Workday ERP to act in real time by detecting key business events and automatically initiating corrective-actions[12].

The results from three healthcare institutions show that such integration not only reduces process latency but also enhances data accuracy and workflow coordination across departments. These outcomes demonstrate that AI can significantly improve organizational agility and governance, enabling healthcare systems to maintain continuous operational readiness even in dynamic environments.

# CONCLUSION

This paper presents a Unique AI-enabled orchestration framework designed to strengthen the event-driven capabilities of Workday ERP in healthcare enterprises system. The framework integrates AI algorithms with Workday's Integration Cloud to detect or build key business process, events, predict outcomes, and automatically trigger responsive actions. These findings demonstrate measurable improvements in process accuracy, operational agility, and coordination across financial and supply chain domains. By enabling intelligent, event-based automation, the discussed approach transforms Workday ERP from a reactive transactional(traditional) platform into a proactive(real time), data-driven decision system.

Looking ahead, future research will focus on extending the proposed architecture to multi-cloud and IoT-integrated healthcare environments, where real-time data from medical devices, sensors, and patient systems can further enhance process intelligence. Additionally, the study identifies opportunities to explore federated learning for privacy-preserving AI, allowing distributed healthcare systems to benefit from shared learning without exposing sensitive data. These advancements will help grow Workday ERP into a real adaptive enterprise platform that supports smarter, more compliant, and more resilient healthcare operations and system.

# ETHICAL DECLARATION

**Conflict of interest:** No Clicflick. **Financing:** No Financing

**REFERENCES**

[1] O. Johnson et al., "Implementing an artificial intelligence command centre in the NHS: a mixed-methods study," *Health and Social Care Delivery Research*, p. 1, Oct. 2024, doi: 10.3310/tatm3277.

[2] H. Yang et al., "FinRobot: Generative Business Process AI Agents for Enterprise Resource Planning in Finance," *RePEc: Research Papers in Economics*, Jun. 2025, https://econpapers.repec.org/RePEc:arx:papers:2506.01423

[3] K. Al-Assaf, W. Alzahmi, R. Alshaikh, Z. Bahroun, and V. Ahmed, "The Relative Importance of Key Factors for Integrating Enterprise Resource Planning (ERP) Systems and Performance Management Practices in the UAE Healthcare Sector," *Big Data and Cognitive Computing*, vol. 8, no. 9, p. 122, Sep. 2024, https://doi.org/10.3390/bdcc8090122

[4] H. Xu, K. Niu, T. Lu, and S. Li, "Leveraging artificial intelligence for enhanced risk management in financial services: Current applications and future prospects," *Engineering Science & Technology Journal*, vol. 5, no. 8, p. 2402, Aug. 2024, doi: 10.51594/estj.v5i8.1363.

[5] C. Bialas, D. Bechtsis, E. Aivazidou, C. Achillas, and D. Aidonis, "Digitalization of the Healthcare Supply Chain through the Adoption of Enterprise Resource Planning (ERP) Systems in Hospitals: An Empirical Study on Influencing Factors and Cost Performance," *Sustainability*, vol. 15, no. 4, p. 3163, Feb. 2023, doi: 10.3390/su15043163.

[6] M. S. Uddin et al., "A hybrid reinforcement learning and knowledge graph framework for financial risk optimization in healthcare systems," *Scientific Reports*, vol. 15, no. 1, Aug. 2025, doi: 10.1038/s41598-025-14355-8.

[7] R. León and P. L. Sanz, "Modeling Health Data Using Machine Learning Techniques Applied to Financial Management Predictions," *Applied Sciences*, vol. 12, no. 23, p. 12148, Nov. 2022, doi: 10.3390/app122312148.

[8] H. Yang et al., "FinRobot: Generative Business Process AI Agents for Enterprise Resource Planning in Finance," 2025, doi: 10.48550/ARXIV.2506.01423.

[9] J. Trottier, W. V. Woensel, X. Wang, K. Mallur, N. M. El-Gharib, and D. Amyot, "Using Process Mining to Improve Digital Service Delivery," *arXiv (Cornell University)*, Aug. 2024, doi: 10.48550/arxiv.2409.05869.

[10] D. F. Eyo, O. A. Oluwaseun, and K. O. Osobase, "Agentic AI in SAP: Collaborative joule agents across procurement, finance and logistics," *Global Journal of Engineering and Technology Advances*, vol. 24, no. 2, p. 91, Aug. 2025, doi: 10.30574/gjeta.2025.24.2.0236.

[11] M. Benjelloun, H. Hmamed, B. Rzine, and A. Dadda, "Navigating Challenges When Integrating Artificial Intelligence with Enterprise Resource Planning: A Literature Review," *Lecture notes in networks and systems*. Springer International Publishing, p. 562, Jan. 01, 2025. doi: 10.1007/978-3-031-91337-2_53.

[12] M. Dumas et al., "AI-augmented Business Process Management Systems: A Research Manifesto," *ACM Transactions on Management Information Systems*, vol. 14, no. 1, p. 1, Jan. 2023, doi: 10.1145/3576047.